\documentclass{aa}
\usepackage{graphicx}
\usepackage{txfonts}
\usepackage{booktabs,caption}
\usepackage{graphicx}
\usepackage{multicol}
\DeclareMathOperator{\sech}{sech}
\begin{document}

   \title{Flaring stellar disk in the low surface brightness galaxy UGC 7321}
       \titlerunning {Flaring stellar disk in the low surface brightness galaxy UGC 7321}

   \author{S. Sarkar
          \inst{1}
          \and
           C.J. Jog
                \inst{1}
           }

   \institute { Department of Physics,
    Indian Institute of Science, Bangalore 560012, India \\
              \email{suchira@iisc.ac.in}\\
            \email{cjjog@iisc.ac.in}
              }

\abstract
{We theoretically study the vertical structure of the edge-on low surface brightness (LSB) galaxy UGC 7321. This is one of the few well-observed LSBs. We modeled it as a gravitationally coupled disk system of stars and atomic hydrogen gas in the potential of the dark matter halo and treated the realistic case where the rotation velocity varies with radius. We used a dense and compact halo as implied by the observed rotation curve in this model. We calculated the thickness of stellar and HI disks in terms of the half-width at half-maximum of the vertical density distribution in a region of R=0 to 12 kpc using input parameters constrained by observations. We obtain a mildly increasing disk thickness up to R=6 kpc, in a good agreement with the observed trend, and predict a strong flaring beyond this. To obtain this trend, the stellar velocity dispersion has to fall exponentially at a rate of 3.2$\mathrm{R_{D}}$, while the standard value of 2$\mathrm{R_{D}}$ gives a decreasing thickness with radius. Interestingly, both stellar and HI disks show flaring in the outer disk region although they are dynamically dominated by the dark matter halo from the very inner radii. The resulting vertical stellar density distribution cannot be fit by a single $\mathrm{\sech^{2/n}}$ function, in agreement with observations, which show wings at larger distances above the mid-plane. Invoking a double-disk model to explain the vertical structure of LSBs as done in the literature may therefore not be necessary.}


\maketitle

\section{Introduction}
Low surface brightness (LSB) galaxies form a special class of galaxies that lie at the faint end of the galaxy luminosity function. Their central surface brightness $\mu_{B}$ is fainter than $23~\mathrm{mag~arcsec^{-2}}$ in the optical B-band, which is about $~30$ times fainter than that of high surface brightness spirals. Their absolute magnitude $\mathrm{M_{B}}$ lies between -17 to -19, which indicates a disk with low surface density (de Blok \& McGaugh 1996; de Blok et al. 2001). Structurally and hence dynamically, they are different from the high surface brightness (HSB) galaxies in several ways. Although they are rich in HI gas content, they have a low star formation rate, have low metallicities and are therefore under-evolved (Bothun et al. 1997). Most importantly, LSBs are dominated by dark matter halo at all radii (Bothun et al. 1997; de Blok \& McGaugh 1997; de Blok et al. 2001) compared to the HSB galaxies, where the dark matter halo becomes dominant only in the outer disk. Within the optical disk, almost 90\%\  of the total galaxy mass is in dark matter, whereas in HSB galaxies, the baryons and dark matter have a comparable mass at these radii (de Blok et al. 2001; Jog 2012). Dark-matter dominance has also been shown  to be the reason for the lack of star formation and for the absence of strong spiral features in LSBs (Jog 2014; Ghosh \& Jog 2014). All these properties distinguish the dynamics and evolution of the LSBs from that of other galaxies.\par

Observationally, it is challenging to study LSBs because even the brightness at their center is fainter than the night-sky background. Nevertheless, in the past two decades, with the advent of wide-field surveys, observations have been able to detect an increasing number of low surface brightness galaxies, which also indicates that they may constitute a large fraction of galaxies in the local Universe. Several studies have therefore been conducted on the surface photometry and kinematics of LSBs (de Blok et al. 1995; Matthews et al. 1999; Du et al. 2015; Pahwa \& Saha 2018; Honey et al.2018), and mass-modeling using rotation curve decomposition methods (de Blok \& McGaugh 1997; Kurapati et al. 2018). However, very few studies have focused on the vertical structure of these galaxies to date (Matthews 2000; Bizyaev \& Kajsin 2004; O' Brien et al. 2010d). The vertical disk structure, including the density profiles and the vertical thickness measurements, can help us to study the stability and star formation history of LSBs, and thus help to distinguish their evolution history from that of HSB galaxies.

\par In this paper, we aim to study the vertical structure of the late-type (Sd) edge-on LSB galaxy, UGC 7321, through theoretical modeling. In a previous work (Sarkar \& Jog 2018) we have studied the detailed vertical density distribution in the Milky Way for a multi-component disk, where the dark matter halo was found to have a significant constraining effect on the vertical stellar disk distribution in the outer disk, and despite this, the disk was found to be flared by a factor of $\sim 4 $ from R=4 to 22 kpc. The LSBs are known to be dominated by dark matter from the innermost radial region, so it would be interesting to study the vertical disk 
distribution in this setting. This was the motivation for the present work. We show that the actual values of disk thickness critically depend on the radial variation of the disk surface density and velocity dispersion and on the values of the dark matter halo parameters. 

We chose UGC 7321 for this detailed study because it is one of the few LSB galaxies whose vertical structure has been observationally studied. UGC 7321 was first studied in detail by Matthews et al. (1999) in optical and near-infrared imaging. Matthews (2000) analyzed the H- and R-band data and the B-R color maps to probe its vertical structure, and thus measured the vertical scale height of the stellar disk at different galactocentric radii and noted a radially increasing scale height, but did not explore this further theoretically. 
With the advent of recent surveys, there is now increasing observational evidence that the thickness of both stellar and HI distributions in the disk increases with radius, and thus both show flaring in the outer disk region of HSB spirals such as the Milky Way (Momany et al. 2006; Lopez-Corredoira \& Molgo 2014; Wang et al. 2018).
The flaring was shown to arise naturally in a theoretical study for the Milky Way (Sarkar \& Jog 2018).
It is therefore important and interesting to determine whether flaring can be a generic result for LSBs as well.

\par With this aim, we solved for the vertical disk distribution for UGC 7321, using the observational data in the literature and using the observed vertical scale height data up to 5.8 kpc as a constraint, to obtain the necessary input parameters for the model. We thus obtained the disk thickness as a function of galactocentric radius from R=0 to 12 kpc. For the sake of completeness, we also calculated the thickness of the HI disk in this radial range. \newline

In an earlier study, Banerjee \& Jog (2013) showed that the dark matter dominance can lead to the superthin nature of this galaxy. However, unlike this
earlier paper, we here considered the more realistic case where the rotation curve varies with radius
while we solved for the vertical structure for any disk component. This is essential because it can significantly affect the disk thickness for a galaxy in a region where the rotation curve is not flat, as we show here. 
Furthermore, we used more physically justified input parameters that are also constrained by the observed data. \medskip 

This paper is organized as follows. The analytical model and its input parameters, as well as the numerical method, are described in detail in Sect.2. The results for our calculation of disk thickness  and its dependence on various dynamical input parameters of the galaxy are presented and explained in Sect.3. Finally, we present conclusions in Sect.4.

\section{Formulation of the problem}

\subsection{Equations for the multi-component disk model}
We used the gravitationally coupled disk model, which consists of stars and atomic hydrogen gas HI disks embedded in the field of the dark matter halo, as developed by Narayan \& Jog(2002b). The molecular gas $\mathrm{H_{2}}$ has negligible surface density in UGC 7321 (Banerjee et al. 2010), and hence is not included in our model. We adopted the galactocentric cylindrical coordinate system ($R, \phi, z$), where $z=0$ denotes the mid-plane. The stellar and HI disks were taken to be coplanar and thin.

The equation of hydrostatic equilibrium for any $ith^{}$ disk component in the $z$ direction is given by (Rohlfs 1977)

\begin{equation}
\frac{\partial }{\partial z}(\rho_{i}\langle(v_{z}^{2})_{i}\rangle)+\rho_{i}\frac{\partial \Phi_{\mathrm{total}}}{\partial z} = 0 \label{eq:1}
,\end{equation}

\noindent where $\rho_{i}$ represents the mass density of stars or gas as a function of $z$. Here $\Phi_{\mathrm{total}}$ is the joint potential of stars, HI and the dark matter halo.
Here $\langle(v_{z}^{2})_{i}\rangle$ denotes the mean square velocity of the ith component along the $z$ direction, and is taken to be constant along $z$, that is, the disk is taken to be isothermal.

This equation is routinely used to study the vertical density distribution of stars in a galactic disk (Spitzer 1942, Bahcall 1984, Kalberla 2003).
Strictly speaking, the Jeans equation along $z$ (Binney \& Tremaine 1987, Eq. 4-29 c) should be used instead of Eq.(1). For any general orientation of the velocity ellipsoid that is pointing toward the galactic center, it can be shown that the two cross terms involving $\langle(v_{R}v_{z})\rangle$ are $\sim(z^{2}/RR_{D})$ times smaller than the terms kept in Eq.(1); see Mihalas \& Routly (1968) and Binney \& Tremaine (1987). For the Milky Way, this ratio is shown to be $< 1 $ \% for $z$ $<1$ kpc (Bahcall 1984), therefore the cross terms can be dropped from the Jeans equation to give Eq.(1). In UGC 7321 also, the observed disk thickness is small, therefore this ratio also turns out to be lower than a few percent. Hence, the cross terms can be dropped, which leads to Eq.(1). We furthermore checked that the resulting values of disk thickness are indeed small (Section 3), which justifies our use of Eq. (1).

\noindent Now, the Poisson equation for this coupled disk plus halo model can be written as
\begin{equation} 
\frac{1}{R}\frac{\partial}{\partial R}\left(R\frac{\partial \Phi_{\mathrm{total}}}{\partial R}\right)+\frac{\partial^{2}\Phi_{\mathrm{total}}}{\partial z^{2}} = 4\pi G\left(\rho_{\mathrm{Stars}}+\rho_{\mathrm{HI}}+\rho_{\mathrm{Halo}}\right) \label{eq:2}
.\end{equation}

Combining the Poisson equation with the equation for vertical hydrostatic equilibrium, we obtain the following equation, which describes the vertical distribution of stars and gas (where $i=\mathrm{stars}, \mathrm{HI}$),
\begin{equation}
\begin{split}
\langle(v^{2}_{z})_{i}\rangle \frac{\partial}{\partial z}\left(\frac{1}{\rho_{i}}\frac{\partial \rho_{i}}{\partial z}\right) & = -4\pi G\left(\rho_{\mathrm{Stars}}+\rho_{\mathrm{HI}}+\rho_{\mathrm{Halo}}\right) \\
                                                                                                                    & +\frac{1}{R}\frac{\partial}{\partial R}\left(v^{2}_{\mathrm{rot}}(R)\right)\label{eq:3}
\end{split}
.\end{equation}

Here, the radial variation of the joint potential is expressed quantitatively through the observed rotation velocity $ v_{\mathrm{rot}}$ as  ${\partial \Phi_{\mathrm{total}}}/{\partial R}={v^{2}_{\mathrm{rot}}(R)}/{R}$. To calculate the radial term in the Poisson equation, we need to know the rotation velocity at each radius. Therefore we fit a curve through the observed data points (Uson \& Matthews 2003) using the principle of polynomial fitting and obtained the best-fit rotation curve (Fig.1). Using this best-fit curve, we determined the radial variation of the rotation velocity and hence the radial term in the Poisson equation (the last term in Eq.(3)) at each radius. The rotation curve clearly shows that it is not flat over most of the
radial range we considered. We note that this approach was developed and used by Banerjee et al.(2011) to study the scale height in dwarf galaxies, but it was not included in the study of UGC 7321 by Banerjee \& Jog (2013).

\begin{figure}
\centering
\includegraphics[height=2.3in,width=3.2in]{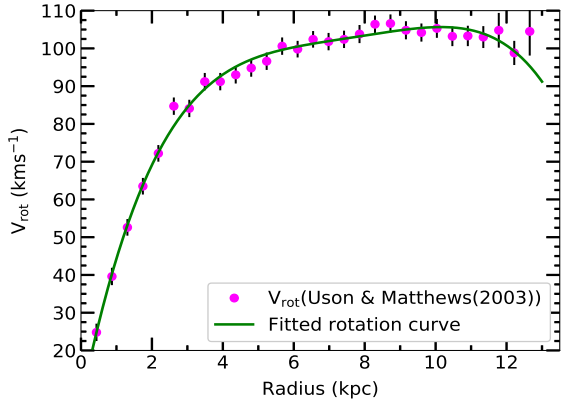}
\caption{Rotation curve for UGC 7321;  rotation velocity vs. radius. The circles represent the observed rotation velocity data, along with the error bars obtained by Uson \& Matthews (2003). The green curve plotted through the data points is the best-fit rotation curve. The curve increases quite steeply nearly up to 4 kpc, then has a slow increment, and becomes almost flat around 10 kpc. At the edge, it slightly declines up to 12 kpc. The observed error bar at each point is also shown. This was provided to us by Matthews (personal correspondence).}
\label{label1}
\end{figure}

\subsection{Input parameters and numerical solution}
To solve the coupled Eq.(3), we need to know the values of the input parameters such as the halo parameters, the vertical velocity dispersion, and the surface density of stars and HI as a function of radius. The surface density values of stars were taken as found from surface photometry by Matthews et al.(1999)  and deduced by Banerjee et al. (2010) with a central value of 50.2 $M_{\sun}~\mathrm{pc^{-2}}$. We note that this is about 12 times lower than the value for the Milky Way (e.g., Narayan \& Jog 2002b). The stellar disk is considered to be razor-thin, and its surface density is taken to fall off exponentially along the radius with radial scale length $R_{D}=2.1~\mathrm{kpc}$ (Matthews et al. 1999). The HI surface density values are from Uson \& Matthews(2003) and are taken from the tabular form of this as provided to us by Matthews (personal communication). These values are given in Table 1. To solve the equation at the observed radial points, that is, at R= 0.0,~0.73, ~1.45, ~2.91, ~4.36, and  ~5.82 in kpc, the values of the HI surface density used are 4.4, ~5.04, ~5.51, ~5.54,~ 4.94, and  ~3.94 in units of $M_{\sun}\mathrm{pc^{-2}}$, respectively, also provided by Matthews (personal communication).

\begin{table}
\caption{Surface density of stars and HI as a function of radius}
\label{table:1}
\centering
\begin{tabular}{l l l}
\hline \hline
Radius & $\Sigma_{\mathrm{stars}}$\tablefootmark{a} & $\Sigma_{\mathrm{HI}}$\tablefootmark{b} \\
(kpc) & $(M_{\sun}\mathrm{pc^{-2}})$ & $(M_{\sun}\mathrm{pc^{-2}})$ \\
 \hline

0.0 & 50.2  & 4.4\\
1.0 & 31.2  & 5.3\\
2.0 & 19.4 & 5.52\\
3.0 & 12.0 & 5.54\\
4.0 & 7.5 &  4.9\\
5.0 & 4.6 &  4.5\\
6.0 & 2.9 &  3.8\\
7.0 & 1.8 &  2.8\\
8.0 & 1.1 &  1.7\\
9.0 & 0.69 &  0.90\\
10.0 & 0.43 & 0.48\\
11.0 & 0.27 & 0.26\\
12.0 & 0.16 & 0.14\\
\hline
\end{tabular}
\tablefoot{
\tablefoottext{a}{Matthews(2000)}\\
\tablefoottext{b}{Uson \& Matthews(2003)}}
\end{table} 

The dark matter halo is taken to be pseudo-isothermal with the density profile given as 
\begin{equation}
\rho(R,z) = \frac{\rho_{0}}{1+\frac{R^{2}+z^{2}}{R^{2}_{c}}}\label{eq:4}
,\end{equation}
\noindent where $\rho_{o}$ and $R_{\mathrm{c}}$ denote the central halo density and the core radius, respectively. Using the corresponding potential ( see Ghosh \& Jog 2014), we derived the halo contribution for the rotation curve (also see Ghosh et al. 2016), and the contribution of the exponential disk was taken from Binney \& Tremaine (1987).
We determined $\rho_{o}$ and $R_{\mathrm{c}}$ 
by fitting the net rotation curve to the observed data points taking the contribution of stars and the dark matter halo and adding them quadratically, as shown in Fig.2. The gas contribution to the rotation curve was found to be negligible for this galaxy (Banerjee et al. 2010) and is therefore not included here. The best-fit to the observed rotation curve gives the resulting halo parameters as $\rho_{0}=0.126\mathrm{M_{\sun}~pc^{-3}} \text{ and } 
R_{\mathrm{c}}= 1.4$~kpc which imply a dense and compact halo.

\begin{figure}
\centering
\includegraphics[height=2.3in,width=3.2in]{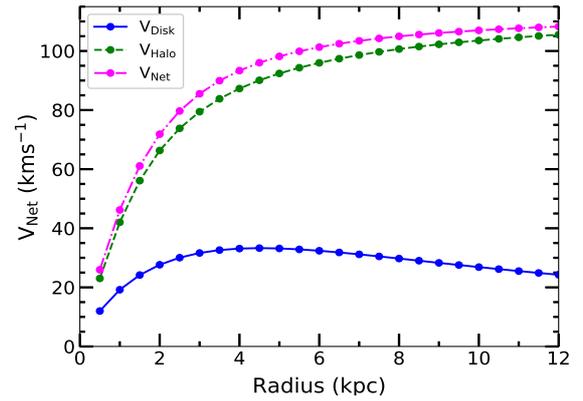}
\caption{This net rotation velocity curve is calculated by adding the contribution of the stellar disk and the dark matter halo quadratically as $\mathrm{V_{Net}}=\mathrm{(V^{2}_{Disk}+V^{2}_{Halo})^{1/2}}$ (where the disk is taken to be razor-thin and has an exponentially falling surface density) so as to obtain the best-fit to the observed rotation velocity data (Uson \& Matthews 2003). This gives the best-fit parameter values for the dark matter halo ($\rho_{0}=0.126\mathrm{M_{\sun}~pc^{-3}},R_{\mathrm{c}}=1.4~\mathrm{kpc})$. This curve clearly shows that the dark matter halo dominates the dynamics of this galaxy at almost all radii.}
\label{label2}
\end{figure}

To explain the scale height of the stellar disk as observed by Matthews(2000), we tried out a set of values for the input parameters. The detailed results of this parameter search are given in Sect.3. For some parameters such as the surface density, the values are fairly well known, and the dark matter halo parameters were determined by fitting the rotation curve as above.  For other parameters we tried a range of values as guided by observations or as
justified on physical grounds, and then chose the values that gave the best-fit to the scale height data.
The details are given in Sects.3.1.1 and 3.1.2. 
All parameters have to be simultaneously varied to obtain the best-fit to the scale height data up to R= 5.8 kpc, therefore we chose to give the sequence of best-fit values for all the parameters in this section. 

\par
The central vertical velocity dispersion of stars is taken to be $20.4 ~\mathrm{km~s^{-1}}$ , as discussed in Matthews (2000). This was not observed directly, but was measured indirectly by a simple analytical model for a single-component disk using the observed central stellar disk scale height. Here, for the best-fit case (found by trial and error), we assumed the velocity dispersion to vary with radius as
\begin{equation}
\sigma_{z}(R) =20.4\exp(-R/3.2R_{D}) \mathrm{km~s^{-1}} \label{eq:5}
.\end{equation}
\noindent Thus the dispersion is taken to fall off exponentially with radius with a scale length  $3.2R_{D}$ instead of $2R_{D}$ , which is typically assumed in literature; for details see Sect.3.1.1. The stellar velocity dispersion is considered to be equal to the HI velocity dispersion from 8 kpc onward because it is found to decrease below the value of $\mathrm{\sigma_{HI}}$ from R=7.2 kpc onward according to Eq.(5). This assumption is justified because stars are formed from the gravitational collapse of gas clouds, hence, they cannot have a velocity dispersion lower than the gas itself. With age, the stellar velocity dispersion is known to only increase; for a detailed discussion, see Sarkar \& Jog (2018). This transition radius depends on the central value of the stellar velocity dispersion and also on the rate at which it falls off with radius and the gas dispersion.

For the gas velocity dispersion we used $\mathrm{\sigma_{HI}}=7\mathrm{km~s^{-1}}$ throughout the disk because this value, corresponding to gas turbulence, was used to obtain the rotation velocity map of this galaxy by Uson \& Matthews (2003).

We stress that to determine the density distribution of any component, Eq.(3) must be solved in a coupled state only because the radial part of the potential has contributions from all the components; otherwise, it may lead to non-convergence in the numerical solution. In particular, we cannot obtain the disk thickness (half-width at half-maximum, HWHM) for stars-alone, for instance, even for the sake of illustration because the dark matter halo gravity on the R.H.S. of Eq.(3) plays a significant role because the galaxy is dominated by the halo gravity at all radii in a typical LSB galaxy. 
This is different for the Galaxy case (which has a flat rotation curve and hence no contribution from radial term), where the disk thickness for stars-alone can be calculated and compared with that in a multi-component disk plus halo case (Sarkar \& Jog 2018).

Using these input parameters, we solved the coupled-disk Eq.(3) numerically using a fourth-order Runge-Kutta method with fifth decimal convergence in the solution. We used the same iterative procedure and boundary conditions as described in Narayan \& Jog (2002b).

\section{Results}
\subsection{Thickness of the stellar disk}
Next we obtained the thickness of the stellar disk using our theoretical model, first at the observed radial points from R=0 to 5.8 kpc to compare them with the observed data and show them in Figures 3 and 4.
As mentioned earlier (Sect.2.2), the observed values were used as a guideline to constrain the input parameters (velocity dispersion of stars and HI),
 along with the other input parameters that are justified on physical grounds,
using which the best-fit values from our model are obtained.
 The HWHM of the vertical density distribution for the best-fit case is taken to denote the disk thickness, as was done in Narayan \& Jog (2002b). Instead of fitting the vertical density distribution to any particular  function, we chose to define the disk thickness in terms of HWHM because the fitting function is not robust and depends on the range $\Delta z$ over which the fitting is done, as we show in Sect.3.4; also see Sarkar \& Jog (2018) for our Galaxy.

To explain the observed light profile in the central region, Matthews (2000) suggested as a possibility the presence of a bulge or more likely a proto-bulge within the central 1 kpc region. In our model we considered UGC 7321 to be a bulgeless disk galaxy, however, as supported in a review by Kautsch (2009). If the effect of bulge were included, it would have changed the resulting disk thickness values within 1 kpc of the center.

Matthews (2000) fitted the observed vertical surface brightness profile at each radius by either an exponential (at the center) or a $\sech(z)$ function (at most other radii) to calculate the scale height values. Interestingly, an increase in scale height by $\sim 40\%$ over radius was noted and was even found to be statistically significant.

\par To compare our results with the observed scale height data $z_{0}$ , we used the conversion relation $\mathrm{HWHM}=\mathrm{z_{0}ln2}$ at the center and $\mathrm{HWHM}=\mathrm{1.32z_{0}}$ at all other radii.
The observed profile at 0.73~kpc was found to lie intermediate between the exponential and $\sech$ by Matthews (2000), who therefore gave the scale height corresponding to both of the fitting functions. For comparison here we have used the value measured assuming a $\sech$ distribution for the sake of consistency in our calculations.

We find that our model is able to produce the mildly radially increasing HWHM values up to$\sim 6$ kpc as observed in the data. Although our HWHM values differ in magnitude from the observed values, the overall trend is the same as found in observations. Interestingly, we find that this observed trend up to 5.8 kpc can only be reproduced when the velocity dispersion values are constrained in a certain way. We explain these findings in the following subsections and show them in Figures 3 and 4. \par

Using the best-fit parameters, as calculated below, we also show the HWHM values from R=0 to 12 kpc at an interval of 1 kpc to study its variation throughout the disk (Fig.5). Our model 
results show a slow increase in HWHM up to $\sim 6$ kpc, as shown in the data, and predicts flaring of the stellar disk beyond that region.

\subsubsection{Dependence on radial variation of the stellar velocity dispersion}
The central vertical velocity dispersion for stars was taken to be $20.4~\mathrm{km~s^{-1}}$, as discussed in Sect.2.2. 
Typically, the vertical velocity dispersion $\mathrm{\sigma_{z}}$ is taken to fall off exponentially with radius with a scale length of $2R_{D}$, as proposed by van der Kruit \& Searle (1981a) to 
explain the constancy of the disk scale height that has been reported in their observations. They suggested that the velocity dispersion of stars should fall off with radius in such a way as to precisely compensate for the decline in the stellar disk surface density and hence its self-gravity. This has become an accepted paradigm although there is no physical justification for this assumption of an exponential fall-off with a scale length of 2$R_{D}$.
The constancy of the scale height itself has been questioned in many papers for external galaxies (de Grijs \& Peletier 1997; Narayan \& Jog 2002a) and for the Milky Way (Kalberla et al. 2014; Sarkar \& Jog 2018). 
When we applied the fall-off in velocity at this rate in our calculation, we found that the thickness (HWHM) decreases with radius, which is exactly opposite to the trend observed in data. We note that the same trend of radially decreasing thickness was also obtained by Banerjee \& Jog (2013), who also used the same assumption of a fall-off with a scale length of $2R_{D}$. 

To resolve this problem, we tried to understand how an increasing disk thickness may arise for a single-component stellar disk. We know that a flat disk can be observed if it satisfies the relation $R_{Vel}=2R_{D}$, where $R_{Vel}$ is the exponential radial scale length of the velocity dispersion $\mathrm{\sigma_{z}}$ of stars and $R_{D}$ is the scale length of the surface density of the radially exponential disk. If the velocity dispersion falls off more slowly than the above required rate, that is, ($R_{Vel}>2R_{D}$), so that the pressure support is higher at a given radius, then it can result in a flared disk. For a single-component disk, an increasing disk thickness can be observed at any value of $R_{Vel}>2R_{D}$. In a gravitationally coupled multi-component system, however, no such simple relation exists between $R_{Vel}$ and $R_{D}$ that would give a disk with constant thickness or a value for which a flared disk would be obtained.
The rate of increase in disk thickness will depend crucially on the value chosen for the ratio $R_{Vel}/R_{D}$. \par

For example, Narayan \& Jog (2002a) investigated the origin of such a radially increasing scale height in two edge-on external galaxies, NGC891 and NGC4565, using the same multi-component disk model. They found this ratio $R_{Vel}/R_{D}$ to be 2.5 and 3, respectively, (or higher) which is able to explain the observed flaring in these two galaxies. 

We therefore have applied this same idea and tried various values of the ratio $R_{Vel}/R_{D}$. We found that an overall flaring in the stellar disk occurs only when $R_{vel}=3.2R_{D}$ or higher, which corresponds to a slower fall-off in velocity dispersion (Fig.3). Although the values of the disk thickness we obtained are higher than the observed values, we note that the correct trend of the overall increase in disk thickness with radius is obtained and thus gives a better overall fit to the data at all radii. Because any value of $R_{vel}>3.2R_{D}$ will give rise to an even higher HWHM, which means a higher mismatch with data, we chose $3.2R_{D}$ as the best-fit case and used it in all further calculations. Our findings once again show that there is no physical justification for taking $R_{vel}=2R_{D}$, because even in the case of a LSB it fails to produce the observed disk thickness. Furthermore, we note that the disk thickness only starts to significantly increase radially when the velocity dispersion is kept constant with radius, that is, beyond 7 kpc (when the stellar velocity dispersion becomes equal to the gas dispersion). \par

We note that the resulting scale height values from our model are consistently higher than the observed values although the radial trend is similar. One reason for this discrepancy could be that the disk thickness sensitively depends on the stellar velocity dispersion and the value of this parameter used in our calculation is model dependent and therefore  not well determined. For example, the estimate of stellar velocity dispersion $\sigma_{z}$ obtained by Matthews (2000) is based on a disk mass estimate from the global stability argument of Efstathiou et al. (1982), and hence, the disk mass used by Matthews (2000) and the resulting velocity dispersion are upper bounds. Any lower value would keep the disk stable against global instability. Thus in reality, if the central disk surface density is lower than that assumed by Matthews (2000), it would give a lower central dispersion. This would give a lower scale height from the model in this paper, which would be in better agreement with observed values. Our results therefore indicate that the true disk density is lower than the upper bound assumed by Matthews (2000). If this is true for other LSBs as well, that is, the disk mass is lower than the upper bound required for global stability, this would result in a more stable disk, which is supported by the observed fact that LSBs lack strong spiral arms. However, we note that in addition to the low mass of the disk, the dominant dark matter halo also plays a major role in preventing local axisymmetric and non-axisymmetric instabilities in LSBs ( Ghosh \& Jog, 2014).\par

Moreover, because it is a low surface brightness galaxy, the observation of the vertical light profile and hence the deduced scale height are limited by the surface brightness limit. Therefore stars at larger vertical distance are likely to be missed in observations. This could therefore lead to an underestimation of the observed scale height and might be another explanation for the discrepancy between our model HWHM values, which are higher than the observed values.

\begin{figure}
\centering
\includegraphics[height=2.3in,width=3.2in]{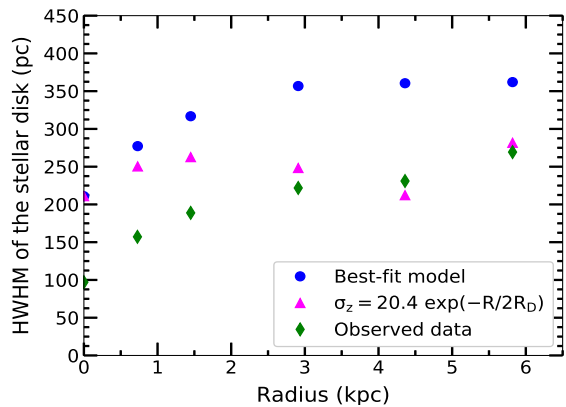}
\caption{Resulting thickness of the stellar disk as a function of radius. When $\mathrm{\sigma_{s}}$ falls off at a rate of $3.2R_{D}$, the HWHM values follow the trend of the given data, while results for an exponential fall-off at a rate of  $\mathrm{2R_{D}}$ give a decrease with radius. From 5.82 kpc it starts to increase because there the velocity dispersion becomes equal to $\mathrm{\sigma_{HI}}$ and is therefore kept constant thereafter.}
\label{label3}
\end{figure} 

\subsubsection{Varying HI velocity dispersion}

\begin{figure}
\centering
\includegraphics[height=2.3in,width=3.2in]{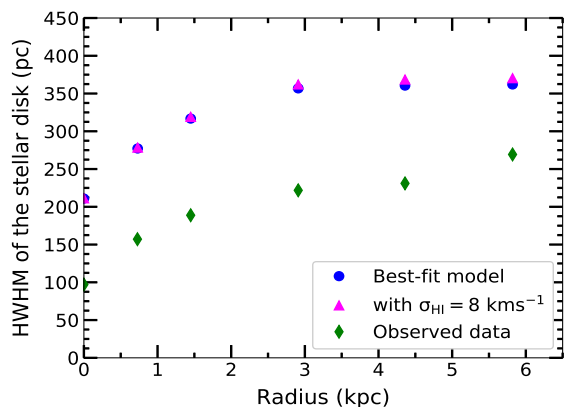}
\caption{Results for variation in stellar disk thickness on changing HI velocity dispersion from 7 to 8 $\mathrm{km~s^{-1}}$. This causes the stellar disk to puff up a little, and therefore the disk thickness is higher than that obtained with 7 $\mathrm{km~s^{-1}}$. Because the values of thickness are closer to the data from $7~\mathrm{km~s^{-1}}$, this is considered to obtain the best-fit model.}
\label{label4}
\end{figure}

\begin{figure}
\centering
\includegraphics[height=2.3in,width=3.2in]{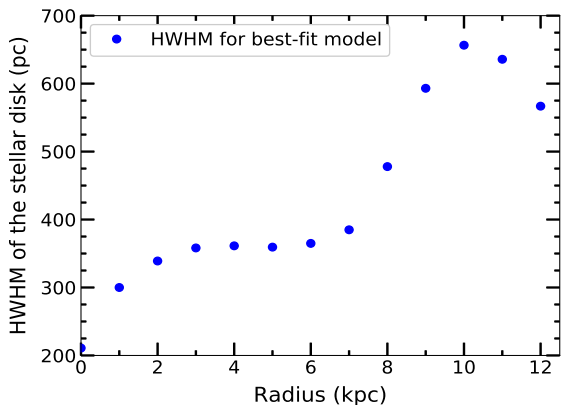}
\caption{Resulting variation of stellar disk thickness (HWHM of the vertical density distribution) as a function of radius from 0 to 12 kpc for our best-fit model. The decline in the rotation velocity at the last two points causes the HWHM to decrease.}
\label{label5}
\end{figure}

As mentioned earlier, we solved for the stellar vertical distribution using the HI velocity dispersion to be $7~\mathrm{km~s^{-1}}$ in the best-fit case. Because the typical observed range of $\mathrm{\sigma_{HI}}$ is $6-9~\mathrm{km~s^{-1}}$ for the outer disk region of galaxies (Kamphuis 1993; Dickey 1996), we also show the stellar disk thickness variation obtained using $\mathrm{\sigma_{HI}}=8~\mathrm{km~s^{-1}}$ for completeness. Fig.4 clearly shows that the HWHM values corresponding to $8~\mathrm{km~s^{-1}}$ are higher than those obtained with $7~\mathrm{km~s^{-1}}$ at all points. The latter choice results in values closer to the observed ones and thus justifies our best-fit choice. Because the velocity dispersion of gas affects the stellar distribution only indirectly in the coupled-disk system (Eq.(3)), the difference in resulting stellar disk thickness values is small.

\subsection{Effect of the radial term in determining the disk thickness}
We have solved the full Poisson equation by including the contribution of the radial term in Eq. (3) (Sect.2.1). As discussed earlier, this term is calculated by fitting the rotation curve to the available data. We note that unlike our Galaxy, the rotation curve here is not flat over most of the radial range considered, with larger gradient toward the inner region. The vertical distribution of stars and HI will therefore be most affected by the radial term in the inner part. \par Up to 10 kpc, the increasing rotation curve causes the radial term to add against the gravity of all the components in the coupled-disk equation, thus causing the disk to puff up by decreasing the effective gravity. This effect becomes more prominent in the inner few kiloparsec region of the disk. 
 \par 
The rotation curve declines near the outermost detected region of the HI disk in UGC 7321, as reported by Uson \& Matthews (2003). Although its origin is far from resolved, some possible interpretations were discussed. A truncation of the mass distribution and hence a probable decline in rotation curve was also discussed in a review by Sofue \& Rubin (2001). We therefore considered this behavior in our calculation to study its possible effect on the disk thickness. We found that the radial term is added to Eq.(3) for R=11 and 12 kpc in such a way as to add to the gravity of all other components. This causes the HWHM to decrease  at these two points. Thus except at the last two points, we found an overall increase in HWHM with radius for both stellar and HI disks that is caused by the inclusion of the radial term.

\subsection{HI disk thickness as a function of radius}

We also determined the HI disk thickness (Fig.6) for this galaxy using the best-fit parameter values that were used to solve for stellar density distribution, which agrees fairly well with the observed values, as discussed below. The intensity maps show a warp, and therefore an analytical form of a flared plus warped model was used in a Gaussian fit to determine the scale heights as a function of radius, owing to the contribution of warp and flare separately (Matthews \& Wood 2003). For the flaring part of the Gaussian, the HWHM was found to increase from $\sim 130~$pc at $R=0$ to $820~$pc at the last measured point of 11.8 kpc, which is in quite good agreement with our result.

Matthews \& Wood (2003) also added a lagging halo to their model to account for the highly extended $z$ emission. Near the mid-plane, however, the warped plus flared model is sufficient to account for the scale heights.

\begin{figure}
\centering
\includegraphics[height=2.3in,width=3.2in]{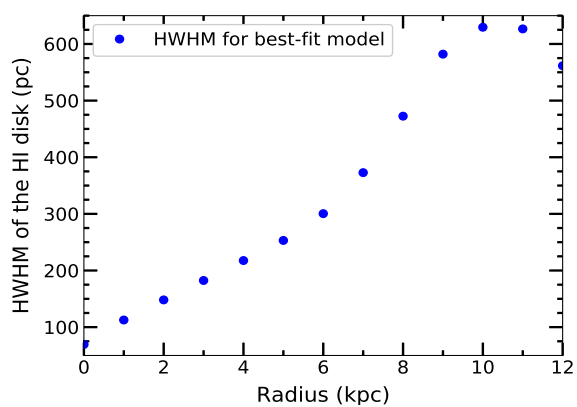}
\caption{Resulting variation in HI disk thickness (HWHM of the vertical density distribution) with radius from R=0 to 12 kpc. The best-fit model parameters taken for stars are used here. The HWHM of the HI disk increases until the drop in rotation velocity at 11 kpc.}
\label{label6}
\end{figure}

\subsection{Determination of the stellar vertical density distribution}

It is well known that the vertical density distribution for a self-consistent, single-component isothermal disk obeys a $\sech^{2}$ distribution law (Spitzer 1942). In a realistic coupled-disk system (as described in Sect.2.1), however, the profile for any disk component would deviate from a simple $\sech^{2}$ behavior. Observationally, it is found to lie between $\sech$ and exponential near the mid-plane, and at larger $z$, it extends to an exponential distribution (van der Kruit 1988). Based on these observed results, van der Kruit(1988) proposed the following general form of the density distribution function to fit along $z:$

\begin{equation}
\rho\left(z\right)=2^{-2/n}\rho_{e}\sech^{2/n}\left(nz/2z_{e}\right).\label{eq:6}
\end{equation}
\noindent Here $n=1,2$ and n$\rightarrow \infty$ correspond to density profiles with $\sech^{2}$, $\sech,$ and an exponential $z$ distribution, respectively.

We note that $2^{-2/n}\rho_{e}$ corresponds to the mid-plane density and $\frac{2}{n}z_{e}$ is used as a measure of scale height. The parameters $2/n$ and $z_{e}$ are obtained with a best-fit analysis of the data. \newline

Although this functional form was not derived on any physical grounds, many authors have studied the observed vertical luminosity profile of a galaxy disk using this form and have given the parameters for the best-fit case (e.g., Bartledrees \& Dettmar 1994; de Grijs et al 1997). Here we first used this function to fit the vertical mass density distribution we obtained from our model at different radii. We calculated the exponent 2/n both as a function of R and $\Delta z$, the vertical interval taken for fitting. \medskip

\begin{figure}
\centering
\includegraphics[height=2.3in,width=3.2in]{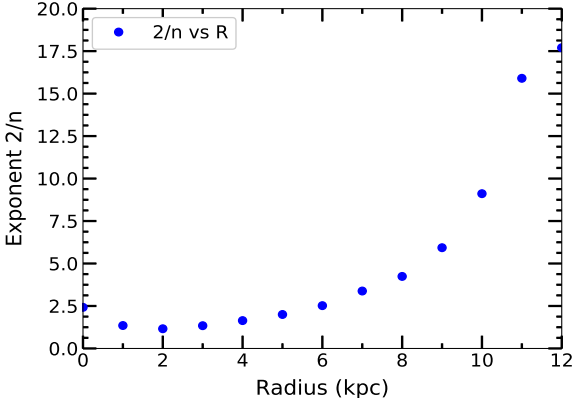}
\caption{Variation in best-fit value of the exponent $2/n$ as a function of radius is obtained when our results for stellar density distribution are fitted with a distribution of type $\sech^{2/n}$ for a range of $\Delta z=200$ pc. The value of $2/n$ is lower than 2 (or $n > 1$) for $R<5$ kpc and greater than 2 (or $n < 1$)
beyond that, i.e., in the outer disk region, which is fully dominated by the dark matter halo.}
\label{label7}
\end{figure}

In Fig.7 we plot the values of 2/n at all galactocentric radii obtained over a $z$ interval of 200 pc, as was done for the Galaxy in a previous study (Sarkar \& Jog 2018). We note that, interestingly, the value of {2/n} starts to be close to 2 or greater than 2, corresponding to $n<1$ from 5 kpc onward, that is, mainly in the outer disk region. This region  is entirely dominated by dark matter halo gravity. A similar trend in results was also noted for our Galaxy (Sarkar \& Jog 2018). We note that this range of $n <1$ is new and
was not considered by van der Kruit (1988).

\begin{figure}
\centering
\includegraphics[height=2.3in,width=3.2in]{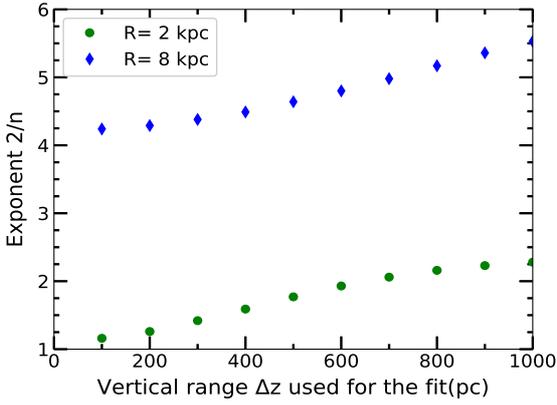}
\caption{Variation in best-fit value of the exponent $2/n$ as a function of fitting range $ \Delta z$ at R=2 and 8 kpc, where our results for vertical stellar density distribution are fit with a distribution of type $sech^{2/n}$. The value of $2/n$ increases with $\Delta z$ regardless of the disk region, in a similar way. Therefore $n$ cannot be regarded as a robust indicator of density profile.}
\label{label8}
\end{figure}

In Fig.8 we show the variation in exponent $2/n$ with vertical range $\Delta z$ that was used to fit $\rho(z)$ at two different regimes of the disk: at an inner radius of 2 kpc, where the rotation curve rises steeply, and at 8 kpc in the outer disk region, where the rotation curve is almost flat. In both cases, the 2/n value increases steadily with increasing $\Delta z$. A similar result was also found for our Galaxy (Sarkar \& Jog 2018). This shows that at any radius in the disk, a fitting function with fixed parameter values cannot provide the best-fit to a $\rho(z)$ distribution over its full range of $z$. Therefore we measured
 the disk thickness in terms of its HWHM instead of $\frac{2}{n}z_{e}$ (see Eq. 6)
, which changes with the value chosen for the range $\Delta z$.

This could be explained as follows: we considered more than one isothermal disk component, for instance, stars and gas, each with a $\sech^{2}$ profile with different surface density and velocity dispersion values. These form a gravitationally coupled system in the field of the dark matter halo which follows a vertical density distribution that is completely different from $\sech^{2}$. Hence the resulting disk distribution cannot be fit by a function with a constant best-fit parameter value throughout the full range of $z$.

We note that Matthews (2000) tried to fit a $\sech(z)$ function over the full vertical range of the observed surface brightness profile at any radius with a constant value of $z_{0}$. The function fits very well in the inner $z$ region, but failed at the profile wings at larger $z$ at almost all radii. In order to explain this, the presence of an additional disk, having a higher scale height, was invoked. Being thicker, this disk was supposed to contribute in the larger $z$ region more than the thinner disk, so as to explain the failure of a single fitting function as given by Eq.(6) at larger $z$. The thick disk was also considered by Matthews (2000) to be a possible reason of the observed increase in scale height with radii. 

However, Fig.8 shows that a physically motivated multi-component disk in a dark matter halo as treated in our model
shows that a single $n$ distribution does not provide a good fit to the vertical stellar density profile at all $z$, as observed.
Our model gives a broader $z$ distribution at high $z$ as observed {\it \textup{without the necessity of invoking a second thicker disk, as done by Matthews (2000)}}. A similar second thicker disk was also invoked to explain the intensity distribution in another LSB galaxy, FGC1540, by Kurapati et al. (2018). Our model shows that the idea of an additional thick disk as invoked in these papers is redundant.

We note that the formation of a thicker disk in a low surface brightness galaxy as proposed in these two papers would be problematic and less likely to occur as such galaxies are usually found in isolated environments (Rosenbaum et al. 2009) and therefore have a low probability of external interaction (Ghosh et al. 2016). Furthermore, they lack molecular clouds and spiral arms (e.g.,Jog 2012; Ghosh \& Jog 2014) , so they do not have an internal source of stellar heating. However, an additional thick disk may yet be viable in some LSBs where its formation could be attributed to mergers, satellite accretion, or radial migration of stars.

\subsection{Resulting luminosity distribution}

Similar to the mass density $\rho(z)$ (van der Kruit, 1988), the general form of the luminosity density distribution is

\begin{equation}
L(R,z)=L_{0}\sech^{2/n}\left(nz/2z_{e}\right)\exp(-R/R_{D})\label{eq:7}
,\end{equation}

\noindent where $L_{0}$ is the luminosity density at the central point, that is, $R=0$ and $z=0$. The vertical surface brightness corresponding to this distribution for an edge-on galaxy can be calculated as 

\begin{equation}
I(R,z)=I(0,0)(R/R_{D})K_1(R/R_{D})\sech^{2/n}\left(nz/2z_{e}\right)\label{eq:8}
,\end{equation}

\noindent where $I(0,0)=2L_{0}R_{D}$ and $K_1$ is the modified Bessel function of the second kind. This equation can be rewritten in terms of the luminosity density $L(R,z)$ and then in terms of the mass density $\rho(R,z)$ as follows:

\begin{align}
I(R,z) & = I(0,0)(R/R_{D})K_{1}(R/R_{D})\left[\frac{L(R,z)}{L_{0}\exp(-R/R_{D})}\right]   \nonumber\\
       & = 2RK_{1}(R/R_{D})\exp(R/R_{D})\left[\frac{\rho(R,z)}{(M_{\odot}/L_{\odot})}\right]. 
\end{align}

In units of magnitude per $\mathrm{arcsecond^{2}}$ , the surface brightness profile can be written as (Binney \& Merrifield 1998)

\begin{equation}
\mu(R,z)=-2.5\mathrm{log_{10}(I/L_{\odot}pc^{-2})+M_{\odot}+21.572}
,\end{equation} \label{eq:10}

\noindent where $M_{\odot}$ denotes the absolute solar magnitude in the observed band. For the H band used by Matthews (2000) 
to observe the vertical surface brightness profiles of UGC 7321, the value of $M_{\odot}$ is $3.32$ (Binney \& Merrifield 1998).

To determine the surface brightness profiles from our calculated data of $\rho(z)$ using Eqs.(9) and (10), we used a mass-to-light ratio $(M_{\odot}/L_{\odot})$ of 1.9 (Bell \& de Jong 2001), which is assumed to be constant at all radii. We plotted all the profiles in Fig.9. The vertical surface brightness profiles clearly flatten with increasing radii because their HWHM increases. In the inner disk region between 2 to 6 kpc, the increase is smaller than that observed at 9 and 10 kpc. At 11 and 12 kpc, the disk thickness decreases as a consequence of the decline  in the rotation curve used in our model, as discussed earlier in Sec. 3.2. Thus, up to 10 kpc in the stellar disk, the surface brightness profiles appear to show flaring, which we have already seen in the measured HWHM.
This again shows that invoking the idea of a second disk
component to explain the observed increase in scale height as done by Matthews (2000) is not needed. A gravitationally coupled multi-component single disk is adequate to explain these observed properties through a physically justified model, as done here.

\begin{figure}
\centering
\includegraphics[height=2.3in,width=3.2in]{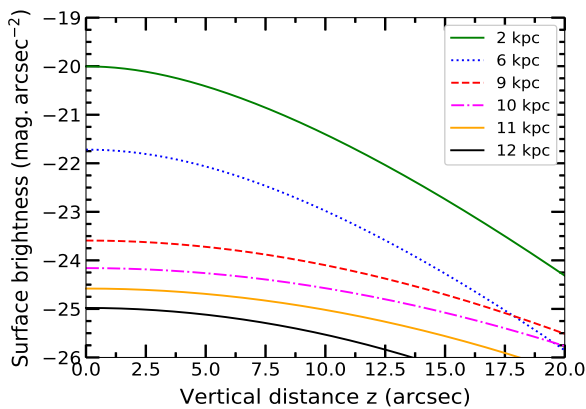}
\caption{Results for the surface brightness distribution along $z$ for R = 2, 6, 9, 10, 11, and 12 kpc. The profiles flatten with increasing radii from R=2 to 10 kpc, indicating an increase in disk thickness and strong flaring at 9 and 10 kpc. The slight decrease in disk thickness seen at 11 and 12 kpc is due to the effect of the observed decline in rotation curve.}
\label{label9}
\end{figure}

In some observations (e.g., van der Kruit \& Searle 1981a), the vertical profiles of the surface brightness at different radii are presented together as a composite $z$-profile by vertically shifting all the profiles to a certain $z$ value. From the slopes of these curves in that composite profile, it is inferred whether there is a significant increase in scale height, for example, a constant scale height is considered when all the curves are identically superimposed on each other and form a single curve. \newline
Here, in figures 10 and 11 we show the same vertical profiles as shown in Fig.9, but vertically shifted to a $z$ value of $5^{''}$ and $15^{''}$ , respectively. We find that the vertical profiles deviate from each other in the figures. This shows that these vertical profiles have different slopes as a result of their different disk thickness, which we have earlier found in our calculation. In these figures, these profiles therefore now show a range of spread in $\mu(z)$ when they are stacked together. The deviation is stronger in the larger and lower $z$ in Fig.10 and 11, respectively, because the points of stacking are different: one at a lower $z$, another at a higher $z$ value. Through these plots, we make a note of caution that the spread in surface brightness can be controlled or even minimized by choosing the point of stacking accordingly. The observed vertical profiles can be stacked together in such a way that can mislead about the concept of a radially constant scale height, which is not true.

\begin{figure}
\centering
\includegraphics[height=2.3in,width=3.2in]{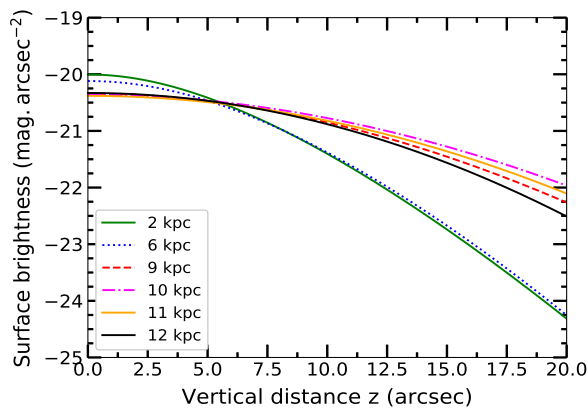}
\caption{Results for the surface brightness distribution along $z$ for R =2, 6, 9, 10, 11, and 12 kpc when the profiles are stacked together at $5^{''}$ to form a composite $z$-profile. The profiles deviate from each other due to different slopes that correspond to different values of disk thickness. The deviation is larger at higher $z$ because the profiles are stacked at a lower $z$ value. If the profiles had the same disk thickness, they would essentially be superimposed on each other and form a single curve.}
\label{label10}
\end{figure}

\begin{figure}
\centering
\includegraphics[height=2.3in,width=3.2in]{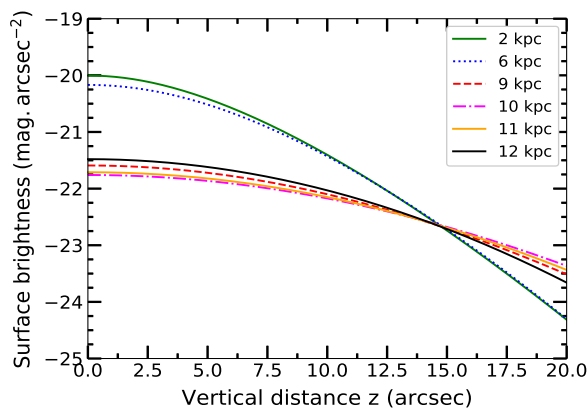}
\caption{Results for the surface brightness distribution along $z$ for R = 2, 6, 9, 10, 11, and  12 kpc when the profiles are stacked together at $15^{''}$ to form a composite $z$-profile. Similar to Fig.10, here also the profiles deviate from each other. The deviation is larger at smaller $z$ because the profiles are stacked at a higher $z$ value. If the profiles had the same disk thickness, they would essentially be superimposed on each other and form a single curve.}
\label{label11}
\end{figure}

\medskip
A similar kind of composite $z$-profile was obtained by Narayan \& Jog (2002a), who studied the vertical structure of two edge-on external galaxies, NGC891 and NGC4565. The spread in surface brightness of around $1~ \mathrm{mag~arcsecond^{-2}}$ in the vertically shifted and stacked $z$-profiles, which is seen in the observed data, could only be reproduced when a varying scale height with radius was considered in the theoretical model instead of a constant scale height, which resulted in a single curve, and this required the velocity to fall off exponentially with a scale length $R_{vel}$ between 2.5 and $3R_{D}$. This shows that any kind of spread observed in the composite $z$-profile of a galaxy could be naturally attributed to varying disk thickness and flaring.

\section{Conclusions}
\medskip

We have studied the vertical disk structure of UGC 7321, a low surface brightness galaxy, by treating the galaxy as a coupled system of stellar and HI disks that are embedded in the dark matter halo. We list our results below:

\noindent 1. Our model can explain the moderate increase in the stellar disk thickness as observed in the data up to $\sim 6$ kpc, predicts even stronger flaring at larger radii up to 12 kpc (up to which the model can be applied because the rotation curve is known up to this radius), and the disk thickness increases by a factor of 3 up to 10 kpc. For completeness, we also studied the HI disk thickness and found a similar kind of flaring throughout the disk. We note that although this galaxy is dominated by a dark matter halo at all radii, the outer disk shows flaring, which indicates that flaring appears to be a generic phenomenon in HSB as well as LSB galaxies.

\noindent 2. Our model HWHM values are higher than the observed values, but they explain the overall flaring trend that is observed. Some possible reasons for this discrepancy, such as the uncertainty in the value of stellar velocity dispersion used as an input parameter, are discussed in Section 3.1.1.

\noindent 3. We showed that the best-fit to the trend in the observed scale height data is obtained when the velocity dispersion falls off exponentially with radius with a scale length of 3.2 $R_{D}$. This corresponds to a much slower fall-off in stellar velocity dispersion than the rate of $2R_{D}$ typically assumed. The latter gives a decreasing disk thickness in the coupled-disk system. This stresses that the assumption  of a fall-off at $2R_{D}$ that is routinely used in the literature is not physically justified in a dynamical modeling of galaxies.

\noindent 4. We showed that taking account of the radial variation in the joint potential for the case of a radially varying rotation curve can affect the disk thickness of both stars and gas to a great extent. This shows that it is in general essential to consider this in dynamical modeling of galaxies with non-flat rotation curves, such as dwarf galaxies.

\noindent 5. The vertical density distribution of stars in LSBs has received very little attention to date. Both the density distribution and the composite luminosity profiles obtained from our model show that a single $sech^{2/n}$ function does not fit the data at all $z$, in agreement with observations (Matthews 2000), which show wings at higher $z$. To explain this behavior in the data, a second thicker disk has been invoked in the past (Matthews 2000; Kurapati et al. 2018). Our model results show that it is not necessary to invoke such a second thicker disk to explain the data.  A gravitationally coupled multi-component single disk in a halo through a physically justified model as done in this paper is adequate to explain the data.

\medskip
We hope that our results, which can predict the nature of the density profiles within and beyond the observed radii of UGC 7321, may motivate observers to study this galaxy and other LSBs in detail. This will help us to understand their dynamics better.

\medskip

\noindent {\bf Acknowledgments: }SS would like to thank CSIR for a fellowship. CJ would like to thank the DST, the Government
of India for support through a J.C. Bose fellowship (SB/S2/JCB-31/2014). We thank Lynn D. Matthews for providing through personal correspondence a tabular form for the data for the rotation curve and the surface density of HI for UGC 7321. We thank the anonymous referee for a constructive and thoughtful report.

\bigskip

\noindent {\bf {References}}
\medskip

\noindent Bahcall J., 1984, ApJ, 276, 156

\noindent Banerjee A., Matthews L. D., \& Jog C. J., 2010, New Astronomy., 15, 89

\noindent Banerjee A. \& Jog C. J., \& Brinks E., Bagetakos I., 2011, MNRAS, 415, 687

\noindent Banerjee, A. \& Jog, C.J. 2013, MNRAS, 431, 582

\noindent Bartledrees, A. \& Dettmar, R.-J. 1994, A\&AS, 103, 475

\noindent Bell, E.F. \&  de Jong, R.S., 2001, ApJ, 550, 212

\noindent Binney, J. \& Tremaine, S. 1987, Galactic Dynamics (Princeton: Princeton Univ. Press)

\noindent Binney, J. \& Merrifield, M. 1998, Galactic Astronomy (Princeton: Princeton Univ. Press)

\noindent Bizyaev D. \&  Kajsin S., 2004, ApJ, 613, 886

\noindent Bothun G., Impey C., \& McGaugh S., 1997, PASP, 109, 745

\noindent de Blok W. J. G., van der Hulst J. M., \& Bothun G. D., 1995, MNRAS, 274, 235

\noindent de Blok W. J. G., \& McGaugh S. S., 1996, ApJ, 469, L89

\noindent de Blok W. J. G., \& McGaugh S. S., 1997, MNRAS, 290, 533

\noindent de Blok W. J. G., McGaugh S. S., \& Rubin V. C., 2001, AJ, 122, 2396

\noindent de Grijs, R., \& Peletier, R.F., 1997, A\&A, 320, L21

\noindent de Grijs, R., Peletier, R.F., van der Kruit, P.C. 1997, A\&A , 327, 966

\noindent Dickey, J.M. 1996, in Unsolved problems of the Milky Way, ed.L. Blitz and P. Teuben (Dordrecht: Kluwer), IAU Symp. 169, 489

\noindent Du W., Wu H., Lam Man I. et al., 2015, AJ, 149, 199

\noindent Ghosh S, \& Jog C.J., 2014, MNRAS, 439, 929

\noindent Ghosh S,  Saini, T.D. \&  Jog C.J., 2016, MNRAS 456, 943

\noindent Honey M., van Driel W., Das M., \& Martin J.-M., 2018, MNRAS, 476, 4488

\noindent Efstathiou G., Lake G., \& Negroponte J., 1982, MNRAS, 199, 1069

\noindent Jog, C.J. 2014, AJ, 147, 132

\noindent Jog C. J., 2012, in Subramaniam A., Anathpindika S., eds, 
ASI Conf. Ser.Vol. 4,  Recent Advances in Star Formation: Observations and Theory.
Astron. Soc. India, Bangalore,  p. 145

\noindent Kalberla P.M.W., 2003, ApJ, 588, 805

\noindent Kalberla P.M.W., Kerp, J., Dedes, L., \& Haud, U. 2014, ApJ, 794, 90   

\noindent Kamphuis J.J.,1993, Ph.D. thesis, University of Groningen

\noindent Kautsch S. J., 2009, PASP, 121, 1297

\noindent Kurapati S., Banerjee A., Chengalur J.N. et al, 2018, MNRAS, 479, 5686

\noindent Lopez-Corredoira, M, \& Molgo, J. 2014, A\&A, 567, A106

\noindent Matthews L. D., Gallagher J. S., \& van Driel W.,  1999, AJ, 118, 2751

\noindent Matthews L. D., 2000, AJ, 120, 1764

\noindent Matthews L. D. \& Wood K., 2003, ApJ, 593, 721

\noindent Mihalas  D. \& Routly P.M, 1968, Galactic Astronomy, A Series of Books in Astronomy and Astrophysics,
San Francisco: W.H. Freeman and Company

\noindent Momany Y., Zaggia. S., Gilmore. G. et al. 2006, A\&A, 451, 515

\noindent Narayan C. A. \& Jog C. J., 2002a, A\&A, 390, L35

\noindent Narayan C. A. \& Jog C. J., 2002b, A\&A, 394, 89

\noindent O' Brien J.C , Freeman K.C, van der Kruit P.C, 2010d, A\& A, 515, A63

\noindent Pahwa I, \& Saha, K., 2018, MNRAS, 478, 4657

\noindent Rohlfs, K., 1977, Lectures on Density Wave Theory, (Berlin:Springer-Verlag)

\noindent Rosenbaum, S. D., Krusch, E., Bomans, D. J., Dettmar, R.-J. 2009, A\& A, 504, 807

\noindent Sarkar S., \& Jog C.J., 2018, A\&A, 617, A142 

\noindent Sofue Y., \& Rubin V., 2001, ARA\&A, 39, 137

\noindent Spitzer, L. 1942, ApJ, 95, 329

\noindent Uson J. M., \& Matthews L. D., 2003, AJ, 125, 2455

\noindent van der Kruit P. C., \& Searle L., 1981a, A\&A, 95, 105

\noindent van der Kruit, P. C. 1988, A\&A, 192, 117

\noindent Wang, H.-F., Liu, C., Xu, Y., Wan, J.-C., Deng, L. 2018, MNRAS, 478, 3367

\end{document}